\documentclass[prl, aps, reprint, superscriptaddress,showpacs,amsmath,amssymb,  floatfix]{revtex4-1}

\usepackage{graphicx}
\usepackage{dcolumn}
\usepackage{bm}
\usepackage{color}
\usepackage{ulem}
\usepackage[utf8]{inputenc}
\usepackage{siunitx}
\usepackage{hyperref}

\definecolor{KommentarPhilip}{gray}{.5}

\newcommand{\nphot}{\bar{n}_{\textrm{r}}} 
\newcommand{\omegac}{\omega_{\textrm{c}}} 
\newcommand{\kappaext}{\kappa_{\textrm{ext}}} 
\newcommand{\kappaint}{\kappa_{\textrm{int}}} 

\newcommand{\nphon}{\bar{n}_{\textrm{m}}} 

\newcommand{\gmV}{g_0} 
\newcommand{\ms}{\gamma} 
\newcommand{\omegap}{\omega_{\textrm{p}}} 

\newcommand{\Omegam}{\Omega_{\textrm{m}}} 
\newcommand{\Omegamod}{\Omega_{\textrm{mod}}} 

\newcommand{\Gammam}{\Gamma_{\textrm{m}}} 

\newcommand{\xzpf}{x_{\textrm{zpm}}} 
\newcommand{\meff}{m_{\textrm{eff}}} 
\newcommand{\Bext}{B_{\textrm{ext}}} 

\newcommand{\PhiZ}{\Phi_\mathrm{0}}

\newcommand{\SUU}{S_{\textrm{UU}}} 
\newcommand{\Sxx}{S_{\textrm{xx}}} 
\newcommand{\SFF}{S_{\textrm{FF}}} 

\newcommand{\LJ}{L_{\textrm{J}}}
\newcommand{\Ic}{I_{\textrm{c}}}
\newcommand{\poc}{\partial \omegac/\partial\Phi}

\newcommand{\Vpiezo}{V_{\mathrm{piezo}}} 

\DeclareSIUnit\fm{\femto\metre}
\DeclareSIUnit\dBm{dBm}
\DeclareSIUnit\mK{mK}
\DeclareSIUnit\pW{pW}
\DeclareSIUnit\kHz{kHz}
\begin{document}

\title{Sideband-resolved resonator electromechanics on the single-photon level based on a nonlinear Josephson inductance}

\author{Philip Schmidt}
\affiliation{Walther-Mei{\ss}ner-Institut, Bayerische Akademie der Wissenschaften, Walther-Mei{\ss}ner-Str. 8, 85748 Garching, Germany}%
\affiliation{Physik-Departement, Technische Universit{\"a}t M{\"unchen}, James-Franck-Str. 1, 85748 Garching, Germany}

\author{Mohammad T. Amawi}%
\affiliation{Walther-Mei{\ss}ner-Institut, Bayerische Akademie der Wissenschaften, Walther-Mei{\ss}ner-Str. 8, 85748 Garching, Germany}%
\affiliation{Physik-Departement, Technische Universit{\"a}t M{\"unchen}, James-Franck-Str. 1, 85748 Garching, Germany}

\author{Stefan Pogorzalek}%
\affiliation{Walther-Mei{\ss}ner-Institut, Bayerische Akademie der Wissenschaften, Walther-Mei{\ss}ner-Str. 8, 85748 Garching, Germany}%
\affiliation{Physik-Departement, Technische Universit{\"a}t M{\"unchen}, James-Franck-Str. 1, 85748 Garching, Germany}

\author{Frank Deppe}%
\affiliation{Walther-Mei{\ss}ner-Institut, Bayerische Akademie der Wissenschaften, Walther-Mei{\ss}ner-Str. 8, 85748 Garching, Germany}%
\affiliation{Physik-Departement, Technische Universit{\"a}t M{\"unchen}, James-Franck-Str. 1, 85748 Garching, Germany}
\affiliation{Munich Center for Quantum Science and Technology (MCQST), Schellingstr. 4, D-80799 München}%

\author{Achim Marx}%
\affiliation{Walther-Mei{\ss}ner-Institut, Bayerische Akademie der Wissenschaften, Walther-Mei{\ss}ner-Str. 8, 85748 Garching, Germany}%

\author{Rudolf Gross}%
\affiliation{Walther-Mei{\ss}ner-Institut, Bayerische Akademie der Wissenschaften, Walther-Mei{\ss}ner-Str. 8, 85748 Garching, Germany}%
\affiliation{Physik-Departement, Technische Universit{\"a}t M{\"unchen}, James-Franck-Str. 1, 85748 Garching, Germany}
\affiliation{Munich Center for Quantum Science and Technology (MCQST), Schellingstr. 4, D-80799 München}%

\author{Hans Huebl}%
\email{huebl@wmi.badw.de}
\affiliation{Walther-Mei{\ss}ner-Institut, Bayerische Akademie der Wissenschaften, Walther-Mei{\ss}ner-Str. 8, 85748 Garching, Germany}%
\affiliation{Physik-Departement, Technische Universit{\"a}t M{\"unchen}, James-Franck-Str. 1, 85748 Garching, Germany}
\affiliation{Munich Center for Quantum Science and Technology (MCQST), Schellingstr. 4, D-80799 München}%

\date{\today}


\begin{abstract}Light-matter interaction in optomechanical systems is the foundation for ultra-sensitive detection schemes \cite{Abbott:2016ki, Moser2013} as well as the generation of phononic and photonic quantum states \cite{Teufel:2011jg, Wollman:2015gx, Lei2016, OckeloenKorppi:2017cx, Aspelmeyer:2014ce, Lu:2015di,Nunnenkamp:2011cp,Qian:2012fw}. Electromechanical systems realize this optomechanical interaction  in the microwave regime. In this context, capacitive coupling arrangements demonstrated interaction rates of up to $\SI{280}{\hertz}$ \cite{Reed2017}. Complementary, early proposals \cite{Blencowe:2007fg, Nation:2008cm,Nation:2016ef, Shevchuk:2017cx} and experiments \cite{Etaki:2008kv, Rodrigues:2019wd} suggest that inductive coupling schemes are tunable and have the potential to reach the vacuum strong-coupling regime. Here, we follow the latter approach by integrating a partly suspended superconducting quantum interference device (SQUID) into a microwave resonator. The mechanical displacement translates into a time varying flux in the SQUID loop, thereby providing an inductive electromechanical coupling. We demonstrate a sideband-resolved electromechanical system with a tunable vacuum coupling rate of up to $\SI{1.62}{\kHz}$, realizing sub-$\mathrm{aN/\sqrt{Hz}}$ force sensitivities. Moreover, we study the frequency splitting of the microwave resonator for large mechanical amplitudes confirming the large coupling. The presented inductive coupling scheme shows the high potential of SQUID-based electromechanics for targeting the full wealth of the intrinsically nonlinear optomechanics Hamiltonian.
\end{abstract}
\maketitle

Designing, investigating, and understanding the optomechanical interaction plays a key role for tailoring the light-matter interaction and hence for testing quantum mechanics \cite{Leggett2002, Poot2012}. In addition, optomechanics is also the basis for detection schemes with extreme sensitivity used in gravitational wave detection \cite{Abbott:2016ki}, mechanical sensing~\cite{Moser2013} as well as the creation of mechanical quantum states~\cite{Wollman:2015gx, Lei2016}. This potential triggered a multitude of realizations of optomechanical systems \cite{Aspelmeyer:2012fy, Aspelmeyer:2014ce} including systems based on superconducting circuits, which define the field of nano-electromechanics \cite{Regal:2008di, Teufel:2011jg,Zhou:2013il, Weber:2014jh, Singh:2014dp}. Here, the photonic cavity is implemented as a microwave resonator and  the electromechanical interaction is typically realized using a mechanically compliant capacitance, which transfers a mechanical displacement into a change of the resonance frequency of the microwave circuit \cite{Regal:2008di, Teufel:2011jg, Zhou:2013il, Weber:2014jh, Singh:2014dp}. It was proposed early on that inductive coupling schemes can allow for higher coupling rates than capacitive ones \cite{Blencowe:2007fg, Nation:2008cm,Nation:2016ef, Shevchuk:2017cx}.
Only  recently, an electromechanical coupling on-par with capacitive approaches was demonstrated by using an inductive coupling scheme based on a lumped-element microwave resonator and a mechanically compliant Josephson inductance \cite{Rodrigues:2019wd}. By design, the linear inductance of the resonator was chosen much larger than the Josephson inductance, limiting the coupling strength. Complementary, it was shown that a distributed resonator with an embedded Josephson inductance allows to design arbitrarily large nonlinearities and can be used to realize two-level systems \cite{Bourassa2012}. Here, we present such an inductive coupling scheme based on a Josephson nonlinearity integrated into a coplanar waveguide resonator (CPW). For this device we find a magnetic field tunable  electromechancial vacuum coupling of up to $\SI{1.62}{kHz}$. Moreover, we find a proportionality factor of $\SI{3.1}{MHz}/\mathrm{T}$, between the coupling rate and the applied magnetic field, exceeding the one reported in Ref.\,\cite{Rodrigues:2019wd} by a factor of 120. In addition, the high coupling rate allows us to detect the mechanical motion with probe powers of $\SI{2.7}{\femto\watt}$ corresponding to an occupation of the microwave resonator with only a few photons.
Our results represent a significant step towards reaching the vacuum strong-coupling regime in electromechanics, allowing the investigation of quantum mechanical effects such as single-photon single-phonon blockade \cite{Aspelmeyer:2014ce, Nunnenkamp:2011cp}, the creation of mechanical quantum states \cite{Lu:2015di}, and the generation of non-classical light~\cite{Qian:2012fw}. 

\begin{figure*}
  \includegraphics[scale=1.2]{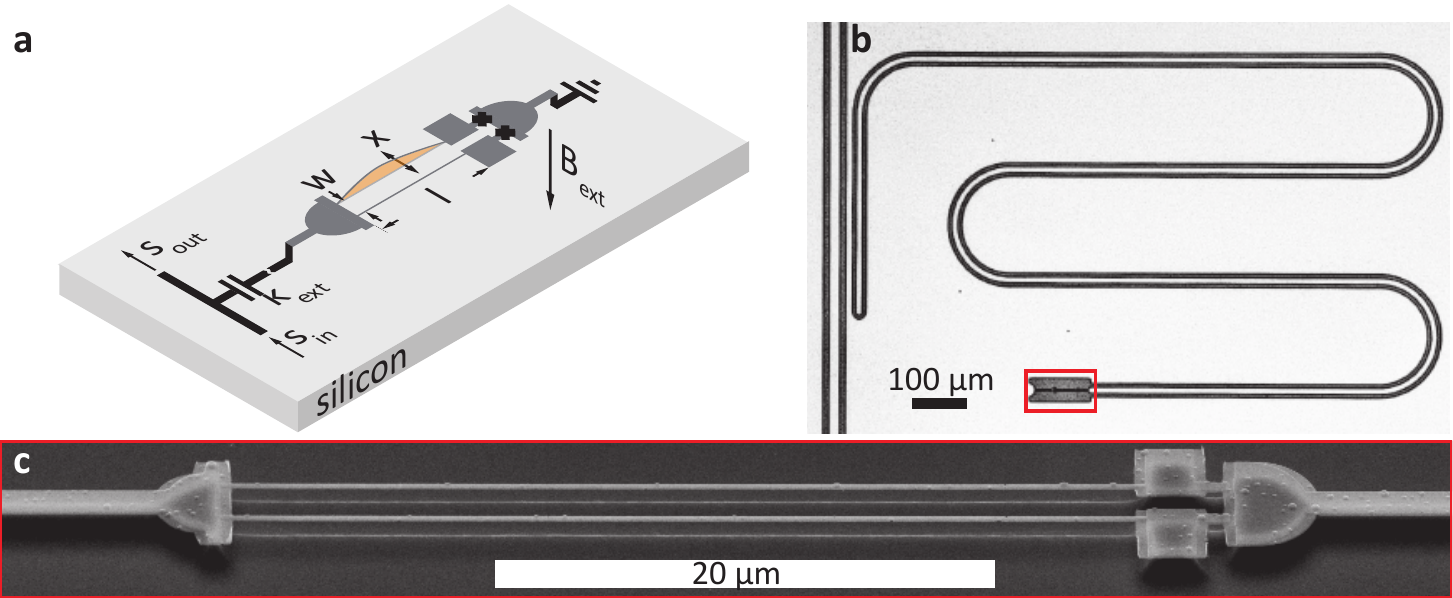}
        \caption{\textbf{Sample schematic and device images.} Panel \textbf{a} shows a schematic of the device. The superconducting $\lambda/4$-coplanar microwave resonator is coupled capacitively with a coupling rate $\kappaext$ to a microwave transmission line acting as feed-line. A dc-SQUID with freely suspended arms shunts the resonator to ground on the other end. The resonance frequency is tunable by varying the flux through the SQUID loop, e.g. by adjusting the external magnetic field $\Bext$. Additionally, the freely suspended SQUID arms (strings) modulate the SQUID inductance with the mechanical frequency via a change in the area of the SQUID loop and hereby realize the electromechanical coupling. Panel \textbf{b} shows a microscope image of the aluminium resonator with the SQUID located at one end fabricated using a lift-off process. Panel \textbf{c} shows a tilted SEM image of a similar partly suspended SQUID structure. Note that the strings investigated in this work have dimensions of $\SI{20}{\micro\metre} \times \SI{110}{\nano\metre} \times \SI{200}{\nano\metre}$.}
        \label{fig:Setup}
\end{figure*}

The nano-electromechanical device discussed here is based on a $\lambda/4$ superconducting CPW resonator which is shunted to ground at one of its ends via a direct-current superconducting quantum interference device (dc-SQUID) (cf.\,Fig.\,\ref{fig:Setup}). As the SQUID acts as a flux-dependent inductance, the resonance frequency of the microwave circuit becomes flux sensitive. In addition, the SQUID loop is partly suspended and contains two nanomechanical string oscillators. Any displacement $\hat{x}$ of the strings translates into a change of the magnetic flux $\Phi$ threading the SQUID loop and, in turn, into the microwave resonance frequency $\omegac$. The mechanically induced frequency shift can be described by the electromechanical interaction Hamiltonian $\hat{H}_\mathrm{int}= \hbar \gmV \hat{a}^\dagger\hat{a}(\hat{b}^\dagger+\hat{b})$, where $\hat{a}$ ($\hat{b}$) are the ladder operators of the microwave resonator (mechanical oscillator). The electromechanical coupling constant \cite{Shevchuk:2017cx,Nation:2008cm,Buks:2007bw,Blencowe:2007fg}
\begin{equation}
\gmV= \frac{\partial\omegac}{\partial \Phi} \delta \Phi = \frac{\partial \omegac}{\partial \Phi} \ms \Bext l \xzpf, 
\label{eq:gom}
\end{equation}
scales with the length $l$ of the string, the zero-point displacement $\xzpf=\sqrt{\hbar / 2 \meff \Omegam}$, and the modeshape $\ms$ of the mechanical resonator. Note, that $\gmV$ is tunable as the applied magnetic field $\Bext$ and the flux to resonance frequency transfer function $\poc$, which we call in the following flux responsivity, can be controlled \textit{in-situ}.

We fabricate the  electromechanical system as an all-aluminium superconducting circuit using electron beam lithography, double-layer shadow evaporation, and reactive ion etching (cf.\,Fig.\,\ref{fig:Setup} and the SI for details). The SQUID contains two freely suspended mechanical string oscillators as well as the two Josephson junctions. Each string has a length $l = \SI{20}{\micro\metre}$, a width $w=\SI{200}{\nano\metre}$, and a thickness $t=\SI{110}{\nano\metre}$, corresponding to an effective mass of $\meff=\SI{0.6}{pg}$. During the device fabrication, we anneal the entire chip at $\SI{350}{\degreeCelsius}$ resulting in tensile stressed aluminum strings with mechanical resonance frequencies $\Omegam \simeq \SI{6.3}{\mega\hertz}$ at millikelvin temperatures. Thus, we obtain a zero-point fluctuation of $\xzpf=\SI{47}{\fm}$. The microwave resonator is coupled capacitively at one of its ends to a microwave CPW feed-line, enabling the microwave spectroscopy of the device. The other end, is shunted to ground via the SQUID. In this way, we obtain a flux-tunable resonator with large flux responsivity $\poc$, which is essential for the realization of a large electromechanical coupling strength.

\begin{figure*}
  \includegraphics[scale=0.6]{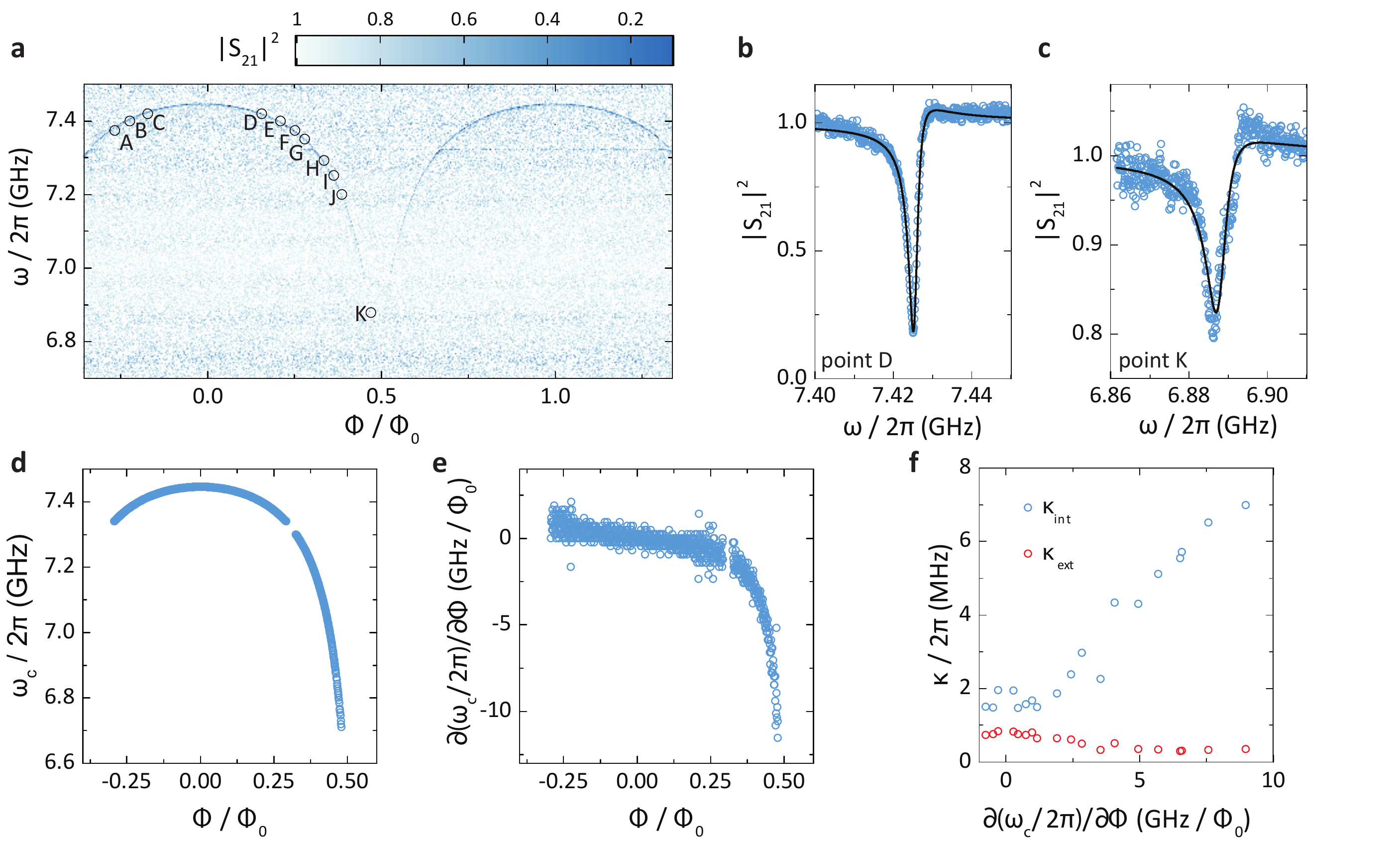}
        \caption{\textbf{Flux tuning of the microwave resonator.} Panel \textbf{a} shows the calibrated microwave transmission as a function of the normalized applied magnetic flux. The tunable resonator exhibits a maximum frequency of $\SI{7.45}{GHz}$ and remains  visible down to $\approx\SI{6.7}{GHz}$. Around $\SI{7.3}{GHz}$, we find a parasitic resonance, which we avoid in our experiments. Additionally, we perform a detailed analysis of the electromechanical coupling rate at the operation points indicated. Panels \textbf{b} and \textbf{c} display the transmission data at points D and K as well as a fit, which allows to quantify the internal and external loss rates of the microwave resonator. Panel \textbf{d} displays the evolution of the microwave resonance frequency as function of the flux bias. The flux to resonance frequency transfer function $\poc$ is computed from this data and shown in panel \textbf{e}, demonstrating responsivities exceeding $\SI{10}{GHz/\PhiZ}$. To judge whether the device operates in the resolved sideband regime, we analyse the transmission data for all operation points and extract the internal and external loss rates as depicted in panel \textbf{f}. While the external coupling rate is rather constant, the internal loss rate increases with increasing responsivity $\poc$.}
        \label{fig:overview}
\end{figure*}

We start the characterization of the microwave circuit by performing microwave transmission measurements as shown in Fig.\,\ref{fig:overview}.
Panel\,\textbf{a} shows the data as a function of the applied flux $\Phi$ (cf.\,SI for details), where the resonator appears as an absorption signature in dark blue. For a quantitative analysis of the evolution of the microwave resonance frequency $\omegac$, we locate the transmission minimum for each flux bias and plot $\omegac$ as well as $\poc$ versus the applied flux in the panels \textbf{d} and \textbf{e}. We find that $\poc$ reaches values of up to $\SI{10}{GHz}/\Phi_0$, underlining the performance of this coupling scheme. In addition, the $\omegac(\Phi)$  dependence allows us to extract the single-junction critical current of the SQUID, $\Ic = \SI{0.44}{\micro\ampere}$, and the minimum Josephson inductance, $\LJ = \SI{0.36}{nH}$, at $\poc =0$ (see also SI). Recording transmission data for selected flux bias points with higher frequency resolution (as shown in panels \textbf{b} and \textbf{c} of Fig.\,\ref{fig:overview}) allows  to determine the external coupling rate $\kappaext$ between the resonator and the feed-line as well as the internal loss rate $\kappaint$ of the microwave circuit. At $\poc=0$, we find a total linewidth of $\kappa/2\pi=(\kappaint+\kappaext)/2\pi=\SI{2.5}{\mega \hertz}$. Panel \textbf{f} shows $\kappaext$ and $\kappaint$ as a function of the flux responsivity. While $\kappaext$ remains nearly constant, $\kappaint$ increases for large $\poc$, which is attributed to the increased sensitivity of the circuit to flux noise. However, even at the bias point K, the device remains in the resolved sideband regime ($\Omegam > \kappa$). 

\begin{figure*}
  \includegraphics[scale=0.6]{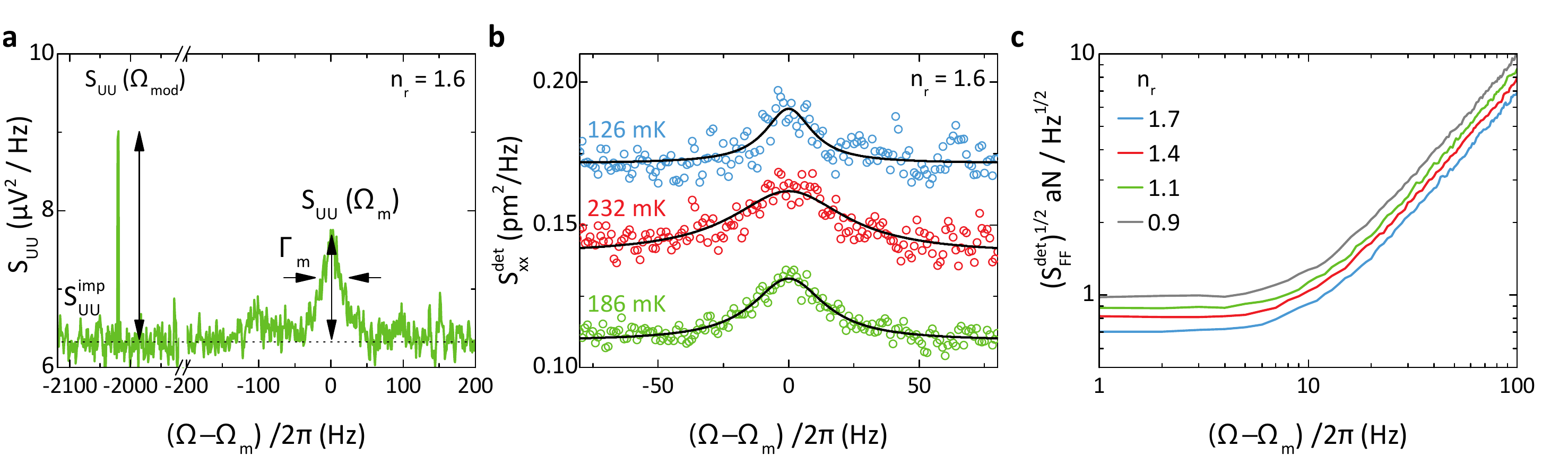}
        \caption{\textbf{Thermal mechanical displacement noise}. Panel \textbf{a} shows the voltage noise spectral density $\SUU$ of the down-converted microwave spectroscopy tone. At $\Omegam/2\pi=\SI{6.34311}{MHz}$, we observe the mechanical signature with a peak amplitude of $\SUU(\Omegam) = \SI{1.26}{\micro\volt^2/Hz}$ and a linewidth of $\Gammam/(2\pi) = \SI{33.6}{Hz}$. In this experiment, we configure the microwave spectroscopy tone to the blue sideband configuration ($\omegap=\omegac+\Omegam$) to enhance the readout efficiency. However, we suppress back-action induced heating by using an ultra-weak probe tone corresponding to an average photon number of $\nphot=1.6$ in the microwave resonator. The sharp peak at $(\Omega-\Omegam)/2\pi \approx \SI{-2}{kHz}$ with an amplitude of $\SUU(\Omegamod) = \SI{2.56}{\micro\volt^2/Hz}$ stems from the phase modulation of the microwave spectroscopy tone ($\phi_0 = 3.94\cdot10^{-4}$). Combining the information of the mechanical signature and the calibration peak, we find an electromechanical vacuum coupling of $\gmV/2 \pi=\SI[separate-uncertainty = true]{1.62(12)}{kHz}$. We further show the mechanical displacement density $\Sxx^{\mathrm{det}}$ in panel \textbf{b} for $T=\SI{126}{mK}$ (blue), $\SI{186}{mK}$ (green), and $\SI{232}{mK}$ (red dots) including the Lorentzian fits to the data (solid lines). The peak area scales as expected with temperature via the mechanical linewidth. Panel \textbf{c} shows the spectral force sensitivity $\SFF^{\mathrm{det}}$ at $T=\SI{126}{mK}$ for the microwave probe powers as indicated. On-resonance, we reach sub-attonewton force sensitivities even for these ultra-low microwave probe powers.
        }
        \label{fig:thermaldisplacement}
\end{figure*}

Next, we investigate the mechanical properties including the electromechanical coupling rate. For this experiment, we probe the microwave sideband fluctuations originating from the interaction of the incident probe tone with the mechanics. The probe tone is set to a blue sideband configuration ($\omegap  = \omegac + \Omegam$) to enable a sensitive detection of the scattered photons~\cite{Aspelmeyer:2014ce}. The resulting signal is down-converted with $\omegap$ and detected using a spectrum analyzer (cf.\,SI). Figure\,\ref{fig:thermaldisplacement}\textbf{a} shows the voltage power spectral density $\SUU(\Omega)$ of this signal for a temperature of $\SI{185}{mK}$, when the microwave resonator is biased to working point K at $\omegac/2\pi = \SI{6.887}{GHz}$ with $\Bext = \SI{-470}{\mu T}$. We find a mechanical signature with a resonance frequency $\Omegam/2\pi=\SI{6.34311}{MHz}$ and a full-width at half-maximum linewidth of $\Gammam/2\pi=\SI{33.6}{Hz}$. In addition, we modulate the frequency of the incident microwave probe tone resulting in a calibration peak visible in panel\,\textbf{a}  at approximately $\SI{-2}{kHz}$. Similar to Refs.\,\cite{Gorodetsky:2010jd,Zhou:2013il}, the comparison of the calibration tone amplitude $\SUU(\Omegamod)$ with the mechanical response $\SUU(\Omegam)\Gammam/2$ allows us to quantify the electromechanical coupling rate $\gmV$. Due to the specific detection scheme used for this experiment, we need to account for an additional factor $\mathcal{Y}$, which relates the transfer function of the mechanical motion to the transmission function of the calibration tone (see SI for details). From this data, we obtain $\gmV/2 \pi=\SI[separate-uncertainty = true]{1.62(12)}{kHz}$. This value exceeds the highest coupling rate of $\SI{280}{\hertz} $\cite{Reed2017} achieved for capacitive coupling by a factor of 5.8. This large coupling strength allows us to use ultra-low probe powers for the detection of the mechanical sidebands. In fact, we use a probe power corresponding to an average photon number of $\nphot = 1.6$ in the microwave resonator for the spectrum shown in Fig.\,\ref{fig:thermaldisplacement}\textbf{a}. At this power level, we suppress back-action induced heating of the mechanical mode. In particular, for the parameters used in our experiment, we estimate an electromechanically induced damping rate of $|\Gamma_{\mathrm{em}}=\SI{-2.6}{Hz}|\ll\Gammam$, resulting in an increase in the thermal phonon occupation by $4\%$ \cite{Aspelmeyer:2014ce}. This contribution is already accounted for in the stated value of $\gmV$ (cf.\,SI). The background of $\SUU(\Omega)$ shows the experimental imprecision noise $\SUU^{\mathrm{imp}}$ of the experiment, which is presently limited by the performance of the cryogenic microwave amplifier used as first amplification stage.

Our calibration technique allows us to express $\SUU(\Omega)$ as the mechanical displacement spectrum $\Sxx(\Omega)$ (cf.\,SI for details). Figure\,\,\ref{fig:thermaldisplacement}\textbf{b} displays these spectra for various temperatures. We observe the expected increase in the phonon number with temperature in the form of an enlarged peak area. This manifests itself as an enhancement of the linewidth $\Gammam$ with temperature (cf.\,SI). This behaviour has been observed for aluminum nanostrings and is attributed to the reduced dimensionality (1D) of the phonon mode \cite{Hoehne2010}. In addition, the imprecision noise floor in these datasets fluctuates, most likely due to variations in resonator and detection performance varying for each temperature.

The detected displacement power spectral density $\Sxx$ directly relates to the force sensitivity $\SFF^{\mathrm{det}}$ via  $\SFF^{\mathrm{det}} = 2 \Sxx(\Omega) / |\chi|^{2}$ \cite{Teufel2009} with the mechanical susceptibility $\chi = [\meff(\Omega^2-\Omegam^2-i\Gammam\Omega)]^{-1}$ (cf.\,SI for details). For $T=\SI{126}{mK}$, we find an on-resonance force sensitivity of $(\SFF^{\mathrm{det}})^{1/2} = \SI{0.70}{aN/\sqrt{Hz}}$ ($\SI{0.98}{aN/\sqrt{Hz}}$) for a probe tone power of $\SI{5.3}{fW}$ ($\SI{2.7}{fW}$) corresponding to $\nphot=1.7\,$ ($0.86$) photons. Here, the large electromechanical coupling rate (cf.\,Fig.\,\ref{fig:thermaldisplacement}\textbf{c}) allows to operate the device at much lower powers compared to capacitive electromechanical devices having demonstrated $\SI{0.54}{aN/\sqrt{Hz}}$ at probe powers of $\SI{1}{pW}$ \cite{Teufel2009}.

\begin{figure*}
  \includegraphics[scale=0.6]{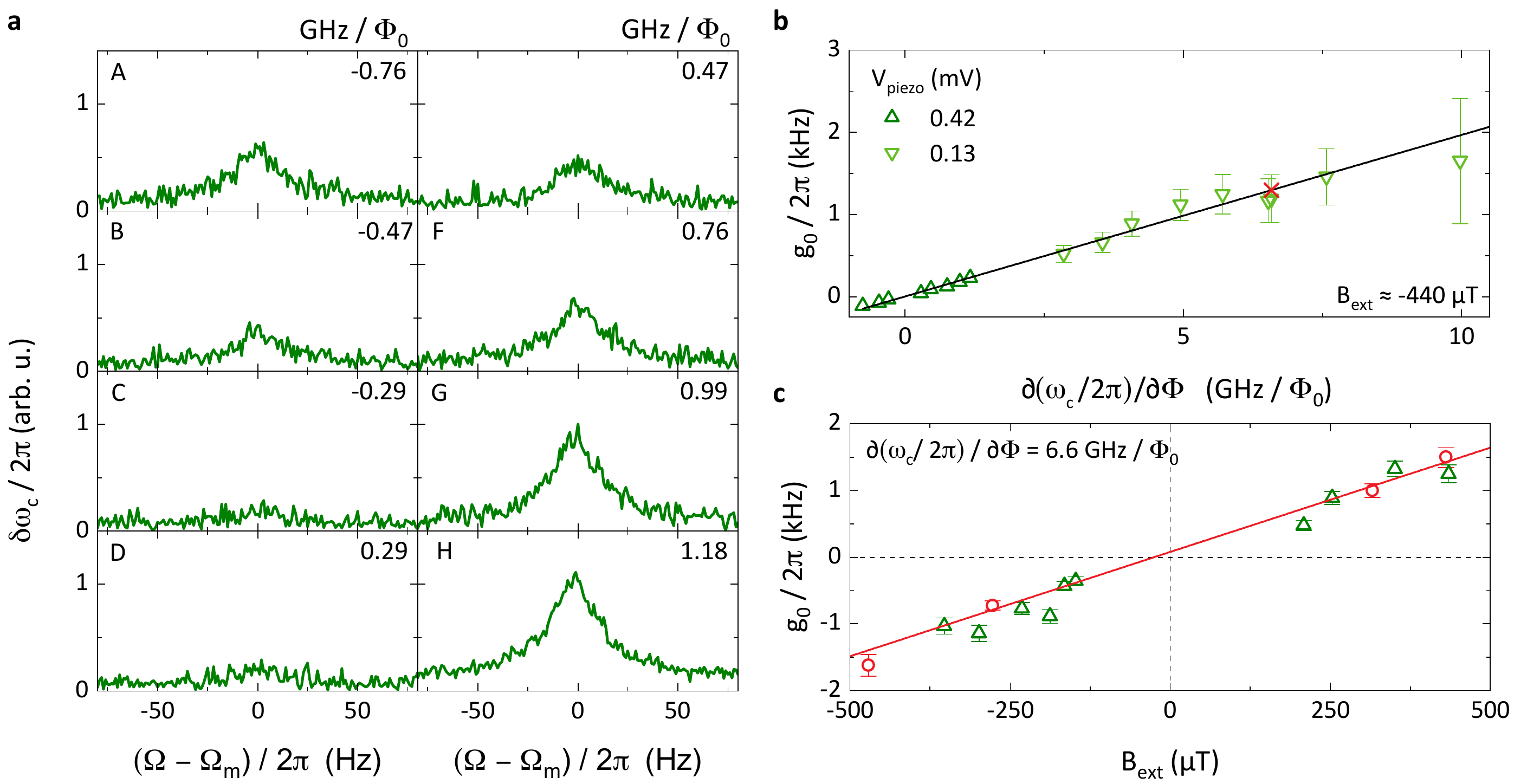}
        \caption{\textbf{Scaling of the electromechanical coupling rate at $\mathbf{T=\SI{126}{mK}}$.} Panel \textbf{a} shows the mechanically driven resonator shift $\delta \omegac$ close to $\Omegam$ for a coherent mechanical drive at angular frequency $\Omega$ generated by a piezo actuator for the various operation points labeled in Fig.\,\ref{fig:overview}. The external static magnetic field is set to $\Bext\approx\SI{-440}{\micro T}$. At magnetic field bias points (e.g. C, D) with low flux responsivity, almost no mechanical signature is observed, in contrast to those points (E to H) with large flux responsivity.  Panel \textbf{b} displays the electromechanical coupling rate $\gmV$ deduced from the response in panel \textbf{a}. The derived values are plotted versus the flux responsivity for low (light green triangles) and high piezo drive amplitudes  (dark green triangles), and are linked to the thermal displacement measurements (red cross). This data confirms the linear scaling of the electromechanical coupling strength with $\poc$. In panel \textbf{c}, we plot the coupling rate $\gmV$ obtained both from driven (green) and thermal displacement measurements (red) versus the applied magnetic field $\Bext$ for a fixed flux responsivity of $\partial (\omegac/2\pi) / \partial \Phi = \SI{6.6}{GHz / \PhiZ}$.
        }
        \label{fig:CouplingLinearities}
\end{figure*}
Next, we investigate the scaling of the electromechanical coupling strength with the applied magnetic field $\Bext$ and the flux responsivity $\poc$, cf.\,Eq.\,(\ref{eq:gom}). To this end, we excite the mechanical motion using an oscillatory mechanical force provided by piezo actuators resulting in a controlled oscillating displacement  of the nano-strings. Under these conditions, the displacement amplitude is much larger compared to the thermal noise driven measurements and hence much easier to detect. Figure\,\ref{fig:CouplingLinearities}\textbf{a} shows the measured resonance frequency shift of the microwave resonator $\delta \omegac$  for various flux responsivities using a coherent excitation force at an approximately constant magnetic field of $\Bext \approx \SI{-440}{\micro \tesla}$. As the frequency shift $\delta\omegac$ is proportional to the product $(\partial\omegac/\partial\Phi) \delta\Phi$,  we observe a large $\delta\omegac$  for large flux responsivity $\poc$. Figure\,\ref{fig:CouplingLinearities}\textbf{b} summarizes the coupling rates derived from the peak amplitudes presented in Fig.\,\ref{fig:CouplingLinearities}\,\textbf{a} (dark green triangles) as well as the results of further experiments with a lower piezo actuator drive power (light green triangles) and links them to displacement noise measurements of $\gmV$ (red cross). The data corroborates the predicted linear scaling of the electromechanical coupling with the flux responsivity of up to $\SI{10}{GHz}/\PhiZ$. 

In a similar fashion, we measure the electromechanical response as a function of the applied magnetic field for a fixed flux responsivity of $\partial (\omega_c/2\pi) / \partial \Phi = \SI{6.6}{GHz/\PhiZ}$ (bias point K) and present the data in Fig.\,\ref{fig:CouplingLinearities}\textbf{c}. Again, we link these data points to the thermal displacement measurements (red circles) and corroborate the linear scaling with $\Bext$.

From Figure\,\ref{fig:CouplingLinearities}\textbf{c}, we extract a scaling factor of $\SI[separate-uncertainty = true]{3.13(20)}{MHz/T}$ for the electromechanical coupling with respect to the applied magnetic field $\Bext$. This value exceeds those of previous reports by two orders of magnitude \cite{Rodrigues:2019wd}. However, we would require $\Bext>\SI{1}{T}$ to reach the vacuum strong coupling regime. This is  not compatible with the present device design. Moving to an in-plane field configuration, in combination with a reduced linewidth of the  microwave resonator, a higher flux responsivity, and superconductors with higher critical fields the vacuum strong coupling regime is in reach. \cite{Shevchuk:2017cx, Nation:2016ef}. Finally, Eq.\,(\ref{eq:gom}) includes a mode-shape factor $\ms$. With our experimental parameters outlined so far, we find $\gamma = 0.99$, which is in good agreement with previous findings \cite{Etaki:2008kv,Rodrigues:2019wd}.

\begin{figure*}
  \includegraphics[scale=0.6]{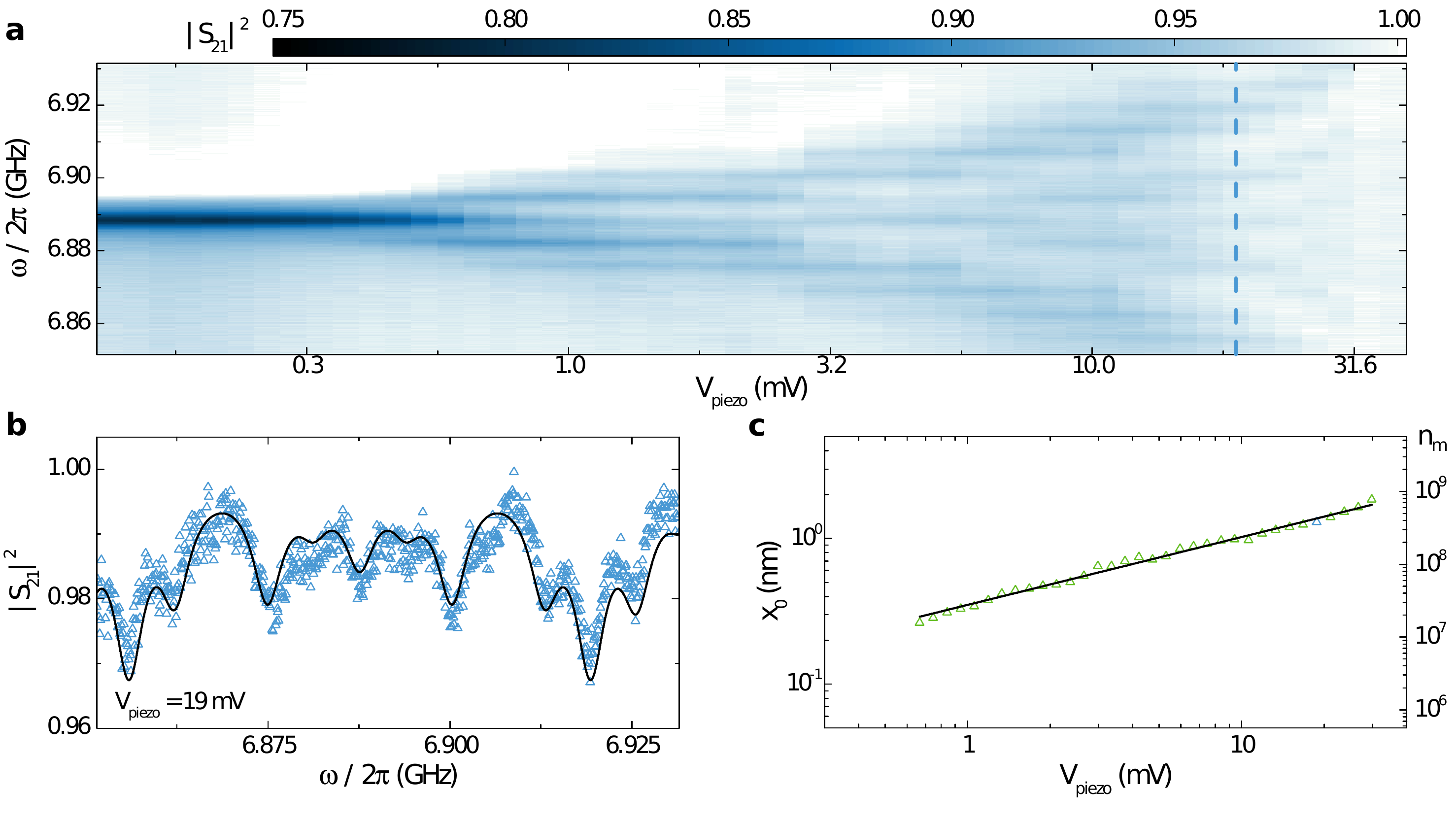}
        \caption{\textbf{Response of the electromechanical system to large amplitudes of the mechanical motion.} In this experiment we excite the mechanical motion resonantly using a piezo actuator. Panel \textbf{a} shows the  transmission of the microwave resonator as a function of the piezo actuator's drive voltage $\Vpiezo$. Above a threshold voltage of approximately $\SI{0.5}{mV}$, the transmission spectrum of the microwave resonator splits into multiple absorption peaks separated by $\Omegam$. To determine the mechanical amplitude $x_0$, we fit Eq.\,(\ref{eq:BesselTrans}) to each transmission spectrum for a fixed $\Vpiezo$. Panel \textbf{b} shows data (blue triangles) and fit (black line) for  $\Vpiezo = \SI{19}{mV}$ demonstrating the good agreement between data and model. Panel \textbf{c} depicts the extracted mechanical amplitude $x_0$ and the corresponding phonon number $\nphon$. We find a scaling of the amplitude $x_0$ with the square root of $\Vpiezo$.}
        \label{fig:Bessel}
\end{figure*}
Next, we utilize the high coupling rate $\gmV$ to study the impact of  large-amplitude displacements on the transmission function of the microwave resonator. In particular, we set the electromechanical coupling rate to $\gmV/2\pi = \SI{1.4}{kHz}$ and drive the mechanical system resonantly using the piezo actuator. For  sufficiently large displacements, we induce frequency shifts of the microwave resonator $\delta \omegac$ comparable or even exceeding the mechanical frequency $\Omegam$. Then, the transmission of the microwave resonator responds by a  splitting of the initial single microwave absorption minimum  as described by~\cite{Schliesser2008, Gorodetsky:2010jd} 
\begin{equation}
    |S_{\mathrm{21}}|^2 = 1-\kappaext\kappa\left(1-\frac{\kappaext}{\kappa}\right)\sum\limits_{n =-\infty}^{\infty} \frac{\left[J_n(\beta)\right]^2}{(\kappa/2)^2 + (\Delta+n\Omegam)}.
    \label{eq:BesselTrans}
\end{equation}
Here, $J_n(\beta)$ is the Bessel function of the first kind and $\beta = \gmV  x_0 / \xzpf \Omegam$. For comparison with the experiment, we consider $|n|\leq 14$. We fit Eq.\,(\ref{eq:BesselTrans}) to the transmission data for each piezo drive voltage $\Vpiezo$, as exemplarily shown in Fig.\,\ref{fig:Bessel}\textbf{b}. This allows us to determine the corresponding coherent mechanical amplitude $x_0$ (cf.\, Fig.\,\ref{fig:Bessel}\textbf{c}). We find that $x_0$ scales with the square root of $\Vpiezo$ indicating that the mechanical displacement entered the nonlinear response regime. Unfortunately, low displacement amplitudes only minimally affect the transmission and hence we do not observe the cross-over between the linear and nonlinear response regime, which would be an independent confirmation of the displacement amplitude \cite{Pernpeintner:2014ks}.

In summary, we present a device implementing an inductive electromechanical coupling scheme based on a superconducting coplanar waveguide resonator and a dc-SQUID operating in the resolved sideband regime. We demonstrate an electromechanical coupling rate of $\SI{1.62}{kHz}$, exceeding values of capacitive coupling schemes by almost an order of magnitude. This high coupling rate enables an ultra-high force sensitivity of $\SI{0.70}{aN/\sqrt{Hz}}$ at an ultra-low  microwave readout power of $\SI{5.4}{fW}$, making this device a very promising low-power force sensor. In addition, the maximum electromechanical coupling of $\SI{1.62}{kHz}$ in combination with the investigated tuning of the electromechanical coupling indicates that this coupling strategy has the potential to reach the vacuum strong coupling regime of electromechanics. Moreover, the tunability of the coupling provides the access to new features of electromechanical systems such as state amplification and readout techniques  using the resonator as a parametric amplifier.

This project is funded from the European Union's Horizon 2020 research and innovation program under grant agreement No 736943 and by the Deutsche Forschungsgemeinschaft (DFG, German Research Foundation) under Germany’s Excellence Strategy – EXC-2111 – 390814868. We gratefully acknowledge valuable scientific discussions with M. Aspelmeyer, K. Fedorov, M. Juan, G. Kirchmair, T. Poeschl, D. Schwienbacher, C. Utschick, S. Weichselbaumer, and D. Zoepfl.


%

\end{document}